\documentclass[reprint,amsmath,amssymb,aps,prl,twocolumn,showpacs]{revtex4}

\usepackage{graphicx}
\usepackage{amsmath,latexsym}
\usepackage{dcolumn}
\usepackage{bm}
\usepackage{sidecap}
\usepackage{multirow}
\usepackage{bm}

\def\bea{\begin{eqnarray}}
\def\nn{\nonumber\\}
\def\eea{\end{eqnarray}}
\def\beq{\begin{equation}}
\def\eeq{\end{equation}}
\def\STO{SrTiO$_3$}
\def\BTO{BaTiO$_3$}
\def\PTO{PbTiO$_3$}

\newcommand{\equ}[1]{Eq.~(\ref{eq:#1})}

\def\P{{\bf P}}
\def\E{{\bf E}}
\def\D{{\bf D}}
\def\rr{{\bf r}}

\def\pa{\partial}

\def\rb{\bar{\rho}}

\def\xh{\hat{x}}

\def\a{\alpha}
\def\b{\beta}
\def\c{\gamma}
\def\d{\delta}

\def\pone{^{(1)}}
\def\ptwo{^{(2)}}
\def\pthr{^{(3)}}

\def\pzex{^{(0,\xh)}}
\def\ponx{^{(1,\xh)}}
\def\ptwx{^{(2,\xh)}}
\def\pthx{^{(3,\xh)}}
\def\pnx{^{(n,\xh)}}
\def\pnpox{^{(n+1,\xh)}}

\def\blib{_{li\b}}
\def\blix{_{lix}}

\def\bzib{_{0i\b}}

\def\bib{_{i\b}}

\def\baib{_{\a,i\b}}

\def\bb{_{\b}}
\def\bab{_{\a\b}}
\def\babc{_{\a\b\c}}
\def\babcd{_{\a\b\c\d}}
\def\bbc{_{\b\c}}
\def\bbcd{_{\b\c\d}}

\def\tix{\tau_{ix}}

\def\veps{\varepsilon}

\def\bVBM{_{\rm VBM}}

\begin{document}

\title{First-principles theory of frozen-ion flexoelectricity}

\author{Jiawang Hong}
\email{hongjw10@physics.rutgers.edu}
\affiliation{ Department of Physics and Astronomy, Rutgers University,
 Piscataway, NJ 08854-8019, USA }

\author{David Vanderbilt}
\affiliation{ Department of Physics and Astronomy, Rutgers University,
 Piscataway, NJ 08854-8019, USA }

\date{\today}

\begin{abstract}

We demonstrate that the frozen-ion contribution to the
flexoelectric coefficient is given solely in terms
of the sum of third moments of the charge density distortions
induced by atomic displacements, even for ferroelectric or
piezoelectric materials.  We introduce several practical
supercell-based methods for calculating these coefficients from
first principles, and demonstrate them by computing the
coefficients for C,
Si, MgO, NaCl, SrTiO$_3$, BaTiO$_3$, and PbTiO$_3$. Three
important subtleties
associated with pseudopotentials, the treatment of surfaces, and
the calculation of transverse components are also discussed.

\end{abstract}


\maketitle


Flexoelectricity (FxE) refers to the linear response of electric
polarization to an applied strain gradient~\cite{Kogan}.  Because
a strain gradient breaks inversion symmetry, FxE is always
symmetry-allowed,
unlike piezoelectricity which arises only in noncentrosymmetric
materials.  The FxE effect is normally negligible on conventional
length scales, but it may become very strong at the nanoscale,
where huge strain gradients can significantly affect
the functional properties of dielectric thin films, superlattices,
and nanostructures. The possibility of large effects at the nanoscale
with application to functional devices has caused a recent explosion of 
experimental interest in flexoelectricity~\cite{Ma-Cross,Cross,
Ma,Catalan-a,Catalan-b,Zubko,Lee}.

There have been remarkably few theoretical studies of FxE,
the main difficulty being that strain gradients are inconsistent
with translational symmetry.  A classical phenomenological theory
focused on lattice-mediated contributions was proposed by
Tagantsev~\cite{tagantsev86,tagantsev91}
and later applied to study FxE properties of dielectrics
by Maranganti and Sharma~\cite{sharma2009}.
A first attempt at a first-principles calculation of FxE is
due to Hong \textit{et al.}~\cite{hong2010}. Recently,
Resta~\cite{resta} developed a first-principles theory of FxE
that was, however, limited to simple elemental insulators such
as Si, and was not implemented in practice.
Thus, unlike piezoelectricity, which is routinely
calculated using modern first-principles methods in a mature theoretical
framework, the theory of FxE remains in a primitive state.

In this Letter, we present a complete theory
of the frozen-ion contributions to the FxE coefficient (FEC),
which were not addressed in Refs.~\cite{tagantsev86,tagantsev91,
sharma2009}.
Working under mixed electric
boundary conditions to be defined shortly, we demonstrate that
the contribution of a given atom to the frozen-ion FEC
is just proportional to the third moment of the change in charge
density induced by its displacement.
This is true for \emph{all} insulating crystals,
from elemental dielectrics to piezoelectrics and ferroelectrics.
Furthermore, we propose several practical supercell-based methods for
extracting the FEC from \textit{ab initio}
calculations, show that these give consistent results, and
discuss their relative advantages.  We report the frozen-ion
FECs for C, Si, MgO, NaCl,
\STO, \BTO, and \PTO, and discuss the trends that emerge from
this data.
Finally, we briefly discuss three important subtleties:
(i) the issue of pseudopotential dependence;
(ii) the question of ``surface contributions'' to the FxE;
and
(iii) the treatment of transverse components using current-density
response.
The extension beyond the frozen-ion case, taking into account
the internal lattice relaxations in response to
strains and strain gradients, will be reported elsewhere.


{\it Theory.}---%
Our approach here is essentially a generalization of the analysis
introduced by Resta~\cite{resta}.
We consider an insulating crystal, fully relaxed
at zero electric field $\E$, and oriented such that one of its primitive
reciprocal lattice vectors lies along $\xh$.
We then identify one entire {\it plane of atoms}, corresponding to
atom $i$ in the home unit cell and its periodic images normal to
$\xh$, and displace the entire plane rigidly by $u\bzib$
in direction $\b$.
This is done under electric boundary
conditions in which the macroscopic $\E$ continues to vanish
away from the displaced plane.  In general this induces
a step in the macroscopic electrostatic potential, so that
if done simultaneously to every $N$'th plane of type
$i$ along $\xh$, it results in an average $E_x\ne0$; instead
what remains unchanged is the electric \emph{displacement} field $D_x$. 
For this reason, we work at ``mixed electric
boundary conditions'' (MEBC) in which we keep the macroscopic
(i.e., supercell-averaged) fields fixed to $E_y=E_z=0$ and
$D_x=4\pi P_{{\rm s},x}$, where $\P_{\rm s}$ is the spontaneous
polarization of the undeformed crystal.

We define the planar-averaged change of charge density induced by this
displacement to be
\beq
f\bib(x)=\frac{\pa\rb(\tau_{ix}+x)}{\pa u\bzib},
\label{eq:f}
\eeq
where $\rb(x)$ is the $y$-$z$ planar average of $\rho(\bf{r})$
and $\bm{\tau}_i$ is the location of atom
$i$ in the unit cell.
We also define the moments of the induced charge redistribution via
\beq
Q\pnx\bib=A\int dx\,f\bib(x)\,x^n,
\label{eq:Q}
\eeq
where $A$ is the cell area normal to $\xh$. Note that the zeroth
moment $Q\pzex\bib$ vanishes due to charge conservation, and that
$Q\ponx\bib$ can be identified as the ``Callen'' or
``longitudinal'' dynamical charge.

By definition the frozen-ion FEC describes the $\P$ induced by
a \textit{homogeneous} strain gradient $\nu$, that is,
\beq
u\blix=\frac{1}{2}\,\nu\,(la+\tau_{ix})^2
\label{eq:HSG}
\eeq
where $l$ is a cell index and $a$ is the lattice constant along
$x$.  In the spirit of Martin~\cite{martin} and Resta~\cite{resta},
we approach this state via
the long-wave ($q\rightarrow0$) limit of a displacement wave
$u\blib=u\bib e^{iq(la+\tix)}$,
where $u\bib=u_\beta$ (independent of $i$) is small enough that a
linear-response approach is appropriate.  Then the charge density
induced by the displacement of sublattice $i$ is
\beq
\rb \bib(x)=u\bib\sum_l e^{iq(la+\tix)}\,f\bib(x-la-\tix).
\eeq
This has Fourier components at $\rb\bib(q+G)$ at all $G=2\pi m/a$, but
we focus on the $G=0$ component defined by
$\rb \bib(q) =(1/a) \int_0^a dx\, e^{-iqx} \rb \bib(x)$ and obtain
\bea
\rb \bib(q) \!&=&\! \frac{u\bib}{a}\int_{-\infty}^\infty dx'\,e^{-iqx'}
     \, f\bib(x') \nn
\!&=&\! \frac{u\bib}{V}\left(-iqQ\bib\ponx -\frac {q^2}{2}Q\bib\ptwx
    +i\frac{q^3}{6}Q\bib\pthx  \right)
\label{eq:Pqx}
\eea
where $x'=x-la-\tix$ is used to obtain
the first line and the series expansion of $e^{-iqx}$ is used to
obtain the second (terms of order $q^4$ and higher have been
dropped), and $V=aA$ is the cell volume~\cite{explan-cell}.
Restoring $u\bib=u_\b$ we get a total
$\rb_\b(q)=\sum_i\rb \bib(q)$,  and using Poisson's equation
in the form $\bar{\rho}(q)=-iqP_x(q)$, this implies
a polarization modulation
\beq
P_{x,\b}(q)=\frac{u_\b}{V}\left(-i\frac{q}{2}Q\bb\ptwx
   -\frac{q^2}{6}Q\bb\pthx \right)
\label{eq:Ptot}
\eeq
where $Q\bb\ptwx=\sum_i Q\bib\ptwx$ and
$Q\bb\pthx=\sum_i Q\bib\pthx$.
The first term of \equ{Pqx} has dropped out due to
the acoustic sum rule $\sum_i Q\bib\ponx=0$.

Now we define the (unsymmetrized) strain tensor and gradient
of the strain tensor to be, respectively,
\beq
\eta\bbc(\rr)=\frac{\pa u_\b(\rr)}{\pa r_\c}, \qquad
\nu\bbcd(\rr)=\frac{\pa \eta\bbc(\rr)}{\pa r_\d}.
\eeq
For the wave $u_\b(\rr)=u_\b e^{iqx}$ this
implies $\eta_{\b x}(q)=iqu_\b$ and $\nu_{\b xx}(q)=-q^2u_\b$,
with other elements such as $\eta_{\b y}$ vanishing.
We also define the (unsymmetrized) frozen-ion piezoelectric
and FxE coefficients to be
\beq
e\babc=\frac{\pa P_\a}{\pa\eta_{\b\c}},\qquad
\mu \babcd=\frac{\pa P_\a}{\pa\nu_{\b\c\d}},
\eeq
which we interpret in the spirit of the long-wave method as
$e\babc=\lim_{q\rightarrow0}\pa P_\a(q)/\pa\eta_{\b\c}(q)$ etc.
Combining the above expressions with \equ{Ptot},
it follows that~\cite{explan-cell}
\beq
e_{x\b x}=-\frac{1}{2V}Q\bb\ptwx,
\label{eq:fpiezo}
\eeq
\beq
\mu_{x\b xx}=\frac{1}{6V}Q\bb\pthx.
\label{eq:fflexo}
\eeq

\equ{fpiezo} expresses the frozen-ion (or ``purely electronic'')
piezoelectric tensor in terms of induced quadrupoles quantified by
the elements of $Q\ptwx$.  This is basically the same as the result
given in the classic paper of Martin \cite{martin}, except that here
all quantities are defined in the MEBC (fixed $D_x$, $E_y$, and
$E_z$).  Similarly, \equ{fflexo} corresponds to the induced-octupole
formulation derived in Resta's Ref.~\cite{resta} and agrees
with Eq.~(22) therein (our $Q\pthr$ is Resta's $AQ\pthr$).
Note, however, that Resta's derivation was limited to elemental (and
therefore non-polar and non-piezoelectric) crystals. Instead,
the derivation here is general, showing that the frozen-ion
FxE response has contributions {\it only} from the
induced octupole term.


\begin{figure}
\includegraphics[width=2.8in]{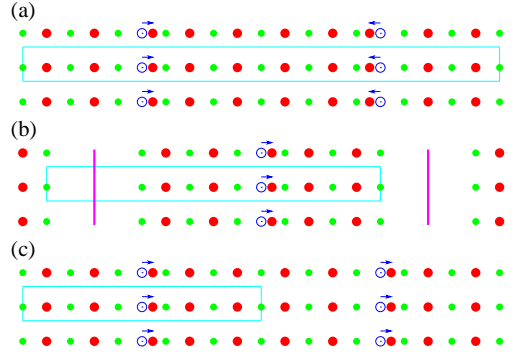}
\caption{\label{fig:model}
(Color online) Supercell geometries. Large (red) and small (green)
dots are two species of atoms; open
dots are atoms before being displaced as shown by arrows.
Rectangles indicate supercells.
(a) Bulk supercell for Method A.
(b) Slab supercell for method B; vertical lines indicate dipole correction
layers in vacuum.
(c) Bulk supercell for Method C.}
\end{figure}

{\it First-principles calculations.}---%
To compute the FECs from
\equ{fflexo} using ab-initio methods, we need to set up
a supercell calculation that allows us to calculate the
$f\bib(x)$ and, from these, the $Q\bib\pthx$, under MEBC
($\Delta D_x$=$E_y$=$E_z$=$0$).
We have designed three independent procedures to accomplish this,
using three different supercell configurations.
In Method A, shown in Fig.~\ref{fig:model}(a),
a supercell is built from $N$ repetitions of the bulk cell, and then
two atomic layers are displaced in {\it opposite} directions
under the usual boundary conditions in which the supercell-averaged
$\E$=0.  Since the induced dipoles are equal and
opposite, they compensate each other, $\Delta\P=\Delta\D=0$,
and the MEBC are satisfied.
In Method B, shown in Fig.~\ref{fig:model}(b), the supercell
contains a slab cut from the bulk material; one central layer
is displaced, and there is an external dipole layer in the vacuum
that is constantly readjusted so that $E_x$ in the vacuum region
does not change.
Again, as long as there is no free charge
on the surfaces, this enforces $\Delta D_x$=0.
Finally, in Method C, illustrated in
Fig.~\ref{fig:model}(c), the supercell is again bulk-like,
but only one layer is displaced, now
using a first-principles code capable of enforcing $\Delta D_x$=0.
In each case, the supercell size or slab thickness has to be
chosen large enough that the induced charge disturbances
$f\bib(x)$ do not overlap or extend to the surface.

The calculations have been performed within density-functional
theory.  We used the local-density approximation~\cite{LDA}
for C, Si, MgO, NaCl and \STO, and the generalized gradient
approximation~\cite{GGA} for \BTO\ and \PTO.
We used SIESTA~\cite{siesta} for
Methods A and B, ABINIT~\cite{Gonze,fixd} for Method C, and
ELK \cite{Elk} for the all-electron calculations to be discussed
later.  Supercells were built from 12 unit cells for the perovskites
and 8 conventional cells for C, Si, MgO and NaCl in Method A
and 4 conventional cells for MgO in Methods B and C; slabs in B are separated by
20\,\AA\ of vacuum.  Atomic displacements of 0.04\,Bohr were used in
SIESTA and ABINIT, and 0.015\,Bohr in ELK.

\begin{table}
\caption{\label{tab:MgO} First and third moments of displacement-induced charge density
for MgO using three different methods.}
\begin{ruledtabular}
\begin{tabular}{ldddddd}
 & \multicolumn{3}{c}{$Q\pone$ ($e$)}
 & \multicolumn{3}{c}{$Q\pthr$ ($e$\,Bohr$^2$)} \\
Method &
\multicolumn{1}{c}{A} &
\multicolumn{1}{c}{B} &
\multicolumn{1}{c}{C} &
\multicolumn{1}{c}{A} &
\multicolumn{1}{c}{B} &
\multicolumn{1}{c}{C} \\
\hline

Mg   & 0.63  & 0.63  & 0.63   &  -8.91  & -8.79  & -8.35  \\
O    & -0.63 & -0.63 & -0.63  & -12.96  & -12.77 & -13.12  \\
Sum   & 0.00 & 0.00 & 0.00&  -21.87 & -21.56 & -21.47 \\

\end{tabular}
\end{ruledtabular}
\end{table}

Table~\ref{tab:MgO} shows the first and third moments of MgO
($Q^{(2)}$=0 by symmetry) from Methods A-C
using identical norm-conserving pseudopotentials.  Clearly the
results are in good agreement,
confirming the consistent implementation of MEBC in all three
approaches.  Methods A and B can be used to calculate FECs using
standard first-principles electronic-structure codes (although
Method B requires a vacuum-dipole capability), but they require
larger supercells.  Converged results can be obtained using smaller
supercells with Method C, but only using a code that implements
fixed-$D$ electric boundary conditions~\cite{fixd}.

Table~\ref{tab:fxc} lists the moments and FECs for several
materials. For elemental and binary dielectrics, it shows that
$|\mu_{xxxx}|$ decreases as ionicity increases.  While
the anion $|Q\pthr|$ increases from MgO to NaCl,
the cation contribution decreases, and cell volume effects
also play an important role.
For all the ABO$_3$ perovskite
structures, the frozen-ion FECs are remarkably similar.
The largest contribution comes from the A atoms, unlike
the (Callen) dynamical charges $Q\pone$, for which Ti and $O_1$
give dominant contributions.
\begin{table}
\caption{\label{tab:fxc}
Lattice constants (of conventional cell~\cite{explan-cell};
$a$ and $c$ for FE \PTO),
first and third moments, and FECs as obtained using Method A.}
\begin{ruledtabular}
\begin{tabular}{ldlddd}
& \multicolumn{1}{c}{$a$} &  &
  \multicolumn{1}{c}{$Q\pone$}  &
  \multicolumn{1}{c}{$Q\pthr$}  &
  \multicolumn{1}{c}{$\mu_{xxxx}$}  \\

& \multicolumn{1}{c}{(Bohr)} &  &
  \multicolumn{1}{c}{($e$)} &
  \multicolumn{1}{c}{($e$\,Bohr$^2$)} &
  \multicolumn{1}{c}{(pC/m)} \\

\hline

 C     & 6.69       & C     & 0.00  & -13.01 & -175.4  \\
 Si    & 10.22      & Si    & 0.00  & -27.94 & -105.7  \\
 MgO   & 7.73       & Mg    & 0.63  & -8.91  & -95.6   \\
       &            & O     & -0.63 & -12.96    \\
                           
 NaCl  & 10.66      & Na    & 0.45  & -1.18  & -47.9    \\
       &            & Cl    & -0.45 & -27.59   \\
                           
 \STO  & 7.31       & Sr    & 0.39  & -54.81 & -144.7  \\
       &            & Ti    & 1.20  & -16.48   \\
       &            & O$_1$ & -0.92 & -27.53   \\
       &            & O$_3$ & -0.33 & -6.59    \\
                   
 \BTO  & 7.52       & Ba    & 0.40  & -65.16 & -141.9  \\
       &            & Ti    & 1.11  & -13.80   \\
       &            & O$_1$ & -0.89 & -27.10   \\
       &            & O$_3$ & -0.31 & -6.78    \\
                   
 \PTO  & 7.43       & Pb    & 0.44  & -59.03 & -156.0  \\
       &            & Ti    & 0.83  & -25.56   \\
       &            & O$_1$ & -0.69 & -23.09   \\
       &            & O$_3$ & -0.29 & -9.57    \\
                   
 \PTO  & 7.35       & Pb    & 0.51  & -57.40 & -148.9  \\
 \;\;(FE)~\cite{note1}      
       & 7.88       & Ti    & 0.76  & -28.41         \\
       &            & O$_1$ & -0.65 & -20.61   \\
       &            & O$_3$ & -0.31 & -9.60    \\

\end{tabular}
\end{ruledtabular}
\end{table}


{\it Rigid-ion model and pseudopotential dependence.}---%
Note that the $Q\pthr$ moments reported in Tables I and II, and
hence the $\mu_{xxxx}$, are all negative.  To see why, consider a
model in which each cation or anion is represented by a spherically
symmetric charge $\rho_i(r)$ that displaces rigidly as a unit.
A brief calculation shows that $Q\pthr_i=\int d^3r\,x^3\,
(-\partial_x \rho_i(r))= 4\pi\int dr\,r^4\,\rho_i(r)$.
The positive nuclear charge at $r$=0 makes no contribution,
so within this model all $Q\pthr_i<0$.  It is not surprising,
then, that the real system shows a similar behavior.

The above analysis also implies that the $Q\pthr_i$, and hence
$\mu_{xxxx}$, should depend on the treatment of the core density
and the pseudopotential construction.
(By contrast, $Q\pone$, and hence $e_{xxx}$, is unaffected.)
For example, if the
ion charge density is partitioned into core and valence contributions
in the above rigid-ion model, both parts will contribute.
We illustrate this in Table~\ref{tab:pseuo} by presenting results for MgO
based on two approaches: an all-electron (AE) calculation, and a
pseudopotential (PS) calculation in which only the change in valence
electron density is used to define $f\bib(x)$, as for
the results presented in Tables \ref{tab:MgO} and \ref{tab:fxc}.
We confirm that AE and PS results agree for the piezoelectric
contributions, but find a significant difference for the FxE ones.

This difference arises as follows.  Suppose the cell-averaged
electrostatic potentials $\bar\phi^{\rm \,AE}$ and $\bar\phi^{\rm \,PS}$
are adjusted such that the valence-band maxima $\veps\bVBM$ agree
between the two bulk calculations.  If the PS is of high quality,
other features of the bandstructure, as well as forces etc., will
show good agreement.  However,  $\bar\phi^{\rm \,AE}\ne\bar\phi^{\rm \,PS}$
because $-e\phi(\rr)$ is typically
much deeper in the AE core region.  Similarly, strain derivatives
will also differ: $d\bar\phi^{\rm \,AE}/d\eta_{xx}\ne
d\bar\phi^{\rm \,PS}/d\eta_{xx}$.  For a strain gradient at
fixed $D_x$ we have $4\pi\Delta P_x=-\Delta E_x=d\bar{\phi}/dx=
(d\bar{\phi}/d\eta_{xx})(d\eta_{xx}/dx)$ so that
$\mu_{xxxx}=(d\bar{\phi}/d\eta_{xx})/4\pi$.
We therefore expect $\mu_{xxxx}^{\rm AE}\ne
\mu_{xxxx}^{\rm PS}$.  Similar considerations apply
to the theory of deformation potentials, which also depend on
the moments $Q\pthr$~\cite{Resta1990,Resta1991}.

The difference between $Q^{(3,{\rm AE})}$ and $Q^{(3,{\rm PS})}$
is unimportant
for some purposes, as for obtaining the spatial gradient of
$\veps\bVBM$ induced by a strain gradient, where it
cancels out of the final result.  Otherwise, there is a simple fix:
for each atom type, we compute a ``rigid core correction'' (RCC)
$Q^{(3,{\rm  RCC})}_i=
4\pi\int dr\,r^4\,[\rho^{\rm AE}_i(r)-\rho^{\rm PS}_i(r)]$
using the densities from free-atom AE and PS calculations, and then
add these $Q^{(3,{\rm  RCC})}_i$ corrections to the
$Q^{(3,{\rm PS})}$ values.
We have done this for Mg and O, obtaining $Q^{(3,{\rm  RCC})}=
-4.85$ and $-0.06$\,$e$\,Bohr$^2$ respectively.  The corrected
values, shown in the last column of
Table \ref{tab:pseuo}, are now in good agreement with the AE ones.

\begin{table}
\caption{\label{tab:pseuo} Moments of MgO obtained from Method
A using all-electron (AE) approach or pseudopotential without
(PS) or with (PS+) rigid-core correction.}
\begin{ruledtabular}
\begin{tabular}{ldddddd}
 & \multicolumn{2}{c}{$Q\pone$ ($e$)}
 && \multicolumn{3}{c}{$Q\pthr$ ($e$\,Bohr$^2$)} \\

& \multicolumn{1}{c}{AE} &
\multicolumn{1}{c}{PS} &&
\multicolumn{1}{c}{AE} &
\multicolumn{1}{c}{PS} &
\multicolumn{1}{c}{PS+} \\

\hline

Mg    & 0.62 & 0.63   & & -14.57 & -8.91 & -13.76 \\
O     & -0.62& -0.63  & & -12.38 & -12.96 & -13.02 \\
Sum$\phantom{aaa}$  & 0.00 & 0.00  & & -26.95 & -21.87 & -26.80 \\

\end{tabular}
\end{ruledtabular}
\end{table}


{\it Surface contributions.}---%
We also considered calculating $\mu_{xxxx}$ by constructing
a slab supercell with two surfaces, as in
Fig.~\ref{fig:model}(b), but applying layer displacements
corresponding to the homogeneous strain gradient of
\equ{HSG}.  Letting $p_x$ be
the total slab (TS) dipole per unit area, we can
define a FEC via $\mu_{xxxx}^{\rm TS}=p_x/\nu L$, where
$\nu=\nu_{xxx}$ and $L$
is the slab thickness.  However, we find that $\mu_{xxxx}^{\rm TS}$
does {\it not} agree with the FEC computed using Methods A-C.
On the other hand, if we compute the FEC from the slope of the
electrostatic potential in the \textit{interior} of the slab using
window convolutions as in Ref.~\cite{resta}, we obtain
$\mu_{xxxx}=-E_x/4\pi\nu$ in good agreement with the results of
Methods A-C.  (In comparison with Method B, however, we
found this method to be more difficult to implement and slower
to converge with slab thickness.)

To explain why $\mu_{xxxx}^{\rm TS}\ne\mu_{xxxx}$, we note that
$\mu_{xxxx}^{\rm TS}$ contains contributions from the slab surfaces.  To see
this, write
$4\pi p_x=\phi_{\rm R}^{\rm vac}-\phi_{\rm L}^{\rm vac}
=\delta\phi_{\rm R} -E_x L -\delta\phi_{\rm L},$
where R and L are right and left surfaces,
and for each surface $\delta\phi=\phi^{\rm vac}-\bar\phi$,
the difference between the vacuum level just outside and the
macroscopic potential just inside the surface.
Dividing by $-4\pi\nu L$, we find
$\mu_{xxxx}^{\rm TS}
 =\mu_{xxxx} + (\delta\phi_{\rm R}-\delta\phi_{\rm L}) /4\pi\nu L$.
Now even if the two surfaces were identical initially, in the presence
of the strain gradient $\nu$ they exist at different strain states,
$\Delta\eta_{xx}=\nu L$, and thus have different $\delta\phi$
values.  In linear response we expect
$\delta\phi_{\rm R}-\delta\phi_{\rm L}
 =\Delta\eta_{xx} (d\delta\phi/d\eta_{xx})$,
from which it follows that
$\mu_{xxxx}^{\rm TS} =\mu_{xxxx} + (d\delta\phi/d\eta_{xx})/4\pi$.
The second term is surface-specific~\cite{explan-surf}
and reflects the dependence of the surface work function on
local strain.

Because we prefer that the FEC should be defined as a
{\it bulk property} independent of surface termination, we
adopt $\mu_{xxxx}$, and not
$\mu_{xxxx}^{\rm TS}$, as our definition of the FEC.
In a sense, $\mu_{xxxx}$ and $\mu_{xxxx}^{\rm TS}$ are analogous
respectively to the ``proper'' and ``improper'' contributions to
piezoelectricity~\cite{vanderbilt-piezo}.


{\it Transverse components.}---%
The derivation of Eqs.~(\ref{eq:fpiezo}-\ref{eq:fflexo}) yielded
$e_{\a\b x}$ and $\mu_{\a\b xx}$ only for the case $\a=x$.  We can
remove this restriction by replacing \equ{f} by
\beq
{\cal P}\baib(x)=\frac{\pa\bar{J_\a}(\tau_{ix}+x)}{\pa \dot{u}\bzib}
\label{eq:Paib}
\eeq
where $\bar{J_\a}(x)$ is the $y$-$z$ planar average of the
current density in direction $\alpha$ induced by the adiabatic motion
$\dot{u}\bzib$ of atomic plane $i$ in direction $\b$, again
under MEBC.  Defining moments
$J\pnx\baib=A\int dx\,{\cal P}\baib(x)\,x^n$, \equ{Ptot} for the
polarization in direction $\alpha$ induced by motions in direction
$\beta$ is replaced by
\beq
P\bab(q)=\frac{u_\b}{V}\left(-iqJ\bab\ponx -\frac{q^2}{2}J\bab\ptwx \right)
\label{eq:PtotJ}
\eeq
where $J\bab\pnx=\sum_i J\baib\pnx$.  It follows that
\beq
e_{\a\b x}=-\frac{1}{V}J\bab\ponx , \qquad
\mu_{\a\b xx}=\frac{1}{2V}J\bab\ptwx .
\label{eq:Jbased}
\eeq
For the longitudinal case $\a$=$x$, this result is equivalent
to Eqs.~(\ref{eq:fpiezo}-\ref{eq:fflexo}), since
continuity implies $\nabla\cdot{\cal P}\bib(\rr)=-f\bib(\rr)$, from
which it follows that $Q\bib\pnpox=(n+1)J_{x,i\b}\pnx$.
By contrast, the moments $J_{\a,i\b}\pnx$ for $\a\!\ne\!x$ contain
additional information about the transverse motions (e.g.,
$J_{y,i\beta}\pzex$ are transverse, or Born, charges).

In principle, the ${\cal P}\baib(x)$ and their moments $J_{\a,i\b}\pnx$ are
computable using the methods of density-functional perturbation
theory.  While we have not implemented such a calculation here,
Eq.~(\ref{eq:Jbased}) formally solves the problem of extending
the present theory to the tensor elements $e_{\a\b x}$ and
$\mu_{\a\b xx}$. By carrying out similar calculations with different
crystal axes aligned along $\xh$,
it should be possible to obtain the full tensors, although care
must be taken to account for the modified interpretation of the
MEBC after the crystal is rotated.


{\it Conclusions.}---%
We have shown that the longitudinal frozen-ion FEC is proportional
to the third moment of induced charge density under MEBC.  An
extension using the second moment of the induced current density
yields also the transverse FECs. This formulation is exact for
\emph{all} insulating crystals. Furthermore, three practical
methods for calculating FECs using \textit{ab initio} methods
have been demonstrated by computing the frozen-ion FECs for
several materials. Issues concerning pseudopotential dependence and
surface effects have also been discussed.
Although it remains to include
lattice contributions associated with internal relaxations that
can occur in response to strains and strain gradients, our work
represents an important step in the direction of a
full first-principles theory of FxE.


This work was supported by ONR grant N00014-05-1-0054.
Computations were done at the Center for Piezoelectrics by Design.



\begin{references}


\bibitem{Kogan} S.M. Kogan, Sov. Phys.-Solid. State {\bf 5}, 2069
  (1964). 

\bibitem{Ma-Cross} W.H. Ma, L.E. Cross, Appl. Phys. Lett. {\bf
 81}, 3440 (2002).

\bibitem{Cross} L.E. Cross, J. Mater. Sci. {\bf 41}, 53 (2006).

\bibitem{Ma} W.H. Ma, Phys. Status Solidi b {\bf 245}, 761 (2008).

\bibitem{Catalan-a} G. Catalan, L.J. Sinnamon and J.M. Gregg, J. 
 Phys.: Condens. Matter. {\bf 16}, 2253 (2004).

\bibitem{Catalan-b} G. Catalan, {\it et al.}, Phys. Rev. B
{\bf 72}, 020102 (2005).

\bibitem{Zubko} P. Zubko, {\it et al.}, Phys. Rev. Lett.  {\bf 99}, (2007).

\bibitem{Lee} D. Lee, {\it et al.}, Phys. Rev. Lett.
{\bf 107}, 057602 (2011).

\bibitem{tagantsev86}  A. K. Tagantsev, Phys. Rev. B {\bf 34},
5883 (1986).

\bibitem{tagantsev91}  A. K. Tagantsev, Phase Transitions {\bf 35},
119 (1991).

\bibitem{sharma2009} R. Maranganti and P. Sharma, Phys. Rev. B
{\bf 80},054109 (2009).

\bibitem{hong2010}J. Hong, G. Catalan, J. F. Scott, and
E. Artacho, J. of Phys.: Condens. Matter. {\bf 22}, 112201
(2010).

\bibitem{resta} R. Resta, Phys. Rev. Lett. {\bf 105}, 127601
(2010).

\bibitem{explan-cell} In Eqs.~(\ref{eq:fpiezo}-\ref{eq:fflexo}),
$V$ may be either the conventional or primitive cell volume, as
long as the sum in $Q\bb\pnx=\sum_i Q\bib\pnx$ runs over
the atoms contained in this volume.

\bibitem{martin} R.M. Martin, Phys. Rev. B {\bf 5}, 1607 (1972).

\bibitem{LDA} J.P. Perdew and A. Zunger, Phys. Rev. B {\bf 23}, 5048 (1981).

\bibitem{GGA} Z. Wu and R. E. Cohen, Phys. Rev. B {\bf 73}, 235116 (2006).

\bibitem{siesta} J. M. Soler, \textit{et al.}, J. Phys.:
Condens. Matter. {\bf 14}, 2745 (2002).

\bibitem{Gonze} X. Gonze, {\it et al.}, Cmp. Mat. Sci. {\bf 25}, 478 (2002).

\bibitem{fixd} J.W. Hong and D.Vanderbilt, arXiv:1106.5668v1

\bibitem{Elk} http://elk.sourceforge.net/

\bibitem{note1} $Q\ptwo$ is 0.22, 0.01, 1.14 and 0.26 for Pb,
Ti, O$_1$ and O$_3$, respectively.

\bibitem{Resta1990} R. Resta, L. Columbo, and S. Baroni,
Phys. Rev. B {\bf 41}, 12358 (1990).

\bibitem{Resta1991} R. Resta, Phys. Rev. B {\bf 44}, 11035 (1991).

\bibitem{explan-surf} While a ``surface contribution'' appears in
Eqs.~(12-13) of Ref.~\cite{tagantsev86}, our
context and our definitions are quite different.

\bibitem{vanderbilt-piezo} D. Vanderbilt, J. Phys. Chem. Solids {\bf 61},
147 (2000).

\end{references}
\end{document}